# Studies on Fabrication of Ag/HgBaCaCuO/CdSe Heterostructures by Pulse-Electrodeposition Route

**D. D. Shivagan, P. M. Shirage and S. H. Pawar***

School of Energy Studies,
Department of Physics,
Shivaji University, Kolhapur- 416 004, INDIA.

*E-mail: pawar_s_h@yahoo.com
shpawar_phy@unishivaji.ac.in

**Abstract :**

Metal∕superconductor∕semiconductor (Ag∕HgBaCaCuO∕CdSe) heterostructures have been successfully fabricated using pulse-electrodeposition technique. The electrochemical parameters are optimized and diffusion free growth of CdSe onto Ag/HgBaCaCuO was obtained by employing under-potential deposition and by studying nucleation and growth mechanism during deposition. The heterostructures are characterized by X-ray diffraction (XRD), full-width at half-maximum (FWHM), scanning electron microscopy (SEM) studies and low temperature four probe electrical resistivity measurements. After the deposition of CdSe the critical transition temperature of HgBaCaCuO films was found be increased from 115 K with $J_c = 1.7 \times 10^3$ A/cm$^2$ to 117.2 K with $J_c = 1.91 \times 10^3$ A/cm$^2$. When the heterostructure was irradiated with red He-Ne laser (2 mW), the $T_c$ was further enhanced to 120.3 K with $J_c = 3.7 \times 10^3$ A/cm$^2$. This increase in superconducting parameters of HgBaCaCuO in Ag/ HgBaCaCuO/CdSe heterostructure has been explained at length in this paper.

**Keywords.** Electrodeposition; Hg-based cuprate; semiconductor; heterostructures; electrical properties.

**PACS Nos** 81.15.Pq; 74.72.Gr; 78.40.Fy; 84.37; 73.40

# 1 Introduction

The higher operating temperatures of high temperature superconductors (HTSC) bring the increased feasibility of hybrid superconductor devices, circuits and systems. Most of the micro cryo-electronics and hybrid devices are based on the use of superconducting thin films integrated with multi-layers of other superconductors, metals, insulators, semiconductors, ferromagnetic materials etc. The important multiplayer structure is an HTSC-insulator-HTSC (SIS) tunnel junction, where it is essential to have HTSC layer maintain its energy gap right upto the interfaces within the SIS sandwich structures. From the material's growth point of view, it is required to avoid both the oxygen deficiency and interdiffusion with the barrier material.

The class of attractive multi-layer structures in the area of superconductivity technology is a hybrid superconductor-semiconductor (super-semi) multi-layers. Super-semi hybrid devices include non-hysteretic Josephson junctions in which the coupling between the HTSC layers is provided by a semiconducting layer and several different types of three terminal devices. These semiconductors can be interestingly tailored in three terminal gated devices like field effect transistors (FET). The coupling strength between the superconductors can be varied by changing the semiconductor doping concentrations, by varying the gate current. At low doping concentrations and low temperatures, the semiconductor becomes an insulator, and the S-Sm-S system behaves as SIS junction, if the semiconductor is thin enough. At higher doping, semiconductor becomes degenerate and acts as a metallic conductor at low temperatures. Hence, the development of tri-layer or bi-layer superconductor-semiconductor heterostructures are important in superconductor technology.

So far, some progress has been made in the growth HTSC thin films on Si [1] and on GaAs [2,3]. To our knowledge, however all successful growths have used intermediate buffer layer between the HTSC films and semiconductor substrates. The major problem in using the semiconductor substrates is the substrate-film inter-diffusion during the in-situ depositions in addition to the formation of micro cracks in the films due to a relatively large difference in thermal expansion coefficients between the semiconductor and HTSC. These fundamental problems in the in-situ growth of the HTSC on Si and GaAs without introducing buffer layers suggest an alternative, and some times preferred, super-semi heterostructures that consists of semiconductor films on top of HTSC.

Heterostructures of BSCCO/GaAs with no buffer layer at the interface were realized in practice by depositing GaAs on top of BSCCO films [4,5], and single crystals [6] using the well established MBE technique and III-V compounds. This showed the great enhancement in $T_c$ from 71 K to 83 K [7]. Rao *et al.* [8] have grown InAs on TlBaCaCuO superconducting films.

The recent experimental evidences [9] showed that photodoping could improve the superconducting transition temperature and the growth of the superconducting phase. In oxygen deficient materials it has been shown that illumination with visible light [10] or ultra-violet light [11], induces persistent photoconductivity [12] and photoinduced superconductivity [13]. A sharp increase in the conductivity of $YBa_2Cu_3O_x$ samples in semiconducting state, has been observed when photon flux from the nitrogen laser exceeding $10^{15}$ photons /cm$^2$, was applied [14].

Further, enhancement in superconducting properties can be possible if photosensitive semiconducting material is deposited onto the superconducting thin films forming super-semi heterojunctions.

The development of such non-diffusive abrupt heterostructures requires suitable growth techniques, which will retain key electronic properties of both super-semi materials. Besides many other thin film deposition techniques, the pulse electrodeposition technique provides a highly reactive mixture on atomic scale which markedly reduces the time and temperature and it gives highly reactive conducting non-porous and fine grained deposits [15]. As this film technique works at room temperature the formation of diffusion during the growth can be avoided only by this technique.

In the present investigation a systematic study has been made to fabricate an Ag/Hg-Ba-Ca-CuO/CdSe heterostructures employing the pulse electrodeposition technique and reported for the first time. Effect on superconducting properties after deposition of CdSe onto Ag/HgBaCaCuO films were studied by changing the geometries of the deposits and also by photo-irradiation and results are discussed in this paper.

## 2. Experimental Procedure:

Cadmium selenide thin films were cathodically electrodeposited from analytical grade 50 mM $CdSO_4$ and 10 mM $SeO_2$ solution onto Ag and Ag/HgBaCaCuO as a substrate. Perkin Elmer potentiostat VersaStat-II with electrochemistry software is used to measure electrochemical characterizations such as linear sweep voltammetry (LSV), Chronoampermetry (CA) properties such as deposition potential, nucleation growth kinetics. Considering this information, the pulse-electrodeposition technique (pulse generator model 1130) was employed to form Ag/HgBaCaCuO/CdSe heterostructure. The X-ray diffraction patterns of these heterostructures are carried out using Philips 3710 diffractogram with $CuK_\alpha$ radiation. The surface morphology of individual deposits and cross section of the heterojunction was carried out using SEM model Philips XL 30 and CAMECA - 30 attached with EDAX. The resistivity was measured using four-probe method where contacts were made by air drying silver paints. The red He-Ne laser with $\lambda = 632.8$ nm, photon energy E = 1.95 eV and power = 2 mW was used to irradiate the samples to study the photoinduced effects on heterostructures.

## 3. Results and Discussion:

In the present investigation the Hg-based oxide superconductor is used because of its highest transition temperature amongst the existing superconductors. We have successfully deposited HgBaCaCuO superconducting films by electrodeposition route at room temperature [16]. The semiconductor material was chosen to be CdSe considering its lattice and thermal agreement and good photosensitive and nano-crystalline properties.

Lattice mismatch between HgBaCaCuO and other semiconductors at 300 K are presented in the following table 1. From the table 1, it is seen that CdSe has lower % lattice mismatch with HgBaCaCuO system compared to other semiconductors. Its band gap is 1.74 eV and the photo-excitations can be carried out in visible spectrum. Varying the ratio of Cd as to Se it can be possible to tailor it as n-type or p-type semiconductors.

### 3.1 Electrochemical Deposition of CdSe on Ag and Ag/HgBaCaCuO

The linear sweep voltammograms (LSV) were recorded for

a) reduction of Cd from 50 mM CdSO$_4$ bath onto Ag substrate,

b) reduction of Se 10 mM SeO$_2$ bath onto Ag substrate

c) deposition of CdSe from (50 mM CdSO$_4$ + 10 mM SeO$_2$) bath onto Ag substrate

d) deposition of CdSe from (50 mM CdSO$_4$ + 10 mM SeO$_2$) bath onto Ag/HgBaCaCuO substrate

and are shown in figure 1 (a-d). The potential range applied was 0 to –1.0 V vs saturated calomel electrode (SCE) for all the measurements. From figure 1 (a) the sudden increase in current at –0.71 V vs SCE can be observed which is due to the reduction of Cd as,

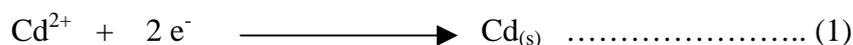

$$Cd^{2+} + 2e^- \longrightarrow Cd_{(s)}$$ ………………….. (1)

The Cd film obtained at –0.8 V vs SCE was found to be blackish gray in color.

Figure 1 (b) shows the LSV for reduction of Se. It is seen that the curve is gradually increasing. The peak at lower potential is due to oxidation process as,

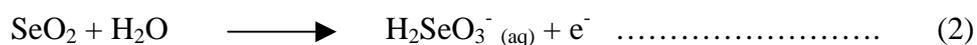

$$SeO_2 + H_2O \longrightarrow H_2SeO_3^-{}_{(aq)} + e^-$$ …………………… (2)

The peak at about –0.22 V vs SCE is due to reduction of Se as [17]

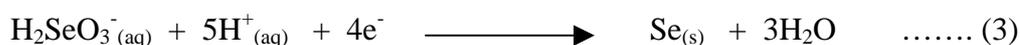

$$H_2SeO_3^-{}_{(aq)} + 5H^+{}_{(aq)} + 4e^- \longrightarrow Se_{(s)} + 3H_2O$$ ……. (3)

Reddish 'Se' films was observed on silver substrate when deposited at –0.5 V vs SCE. Figure 1 (c) shows the sharp increase in current at about –0.85 V vs SCE. This represents the following simultaneous redox reaction. The electrochemical reaction on cathode (Ag) takes place as follows [18].

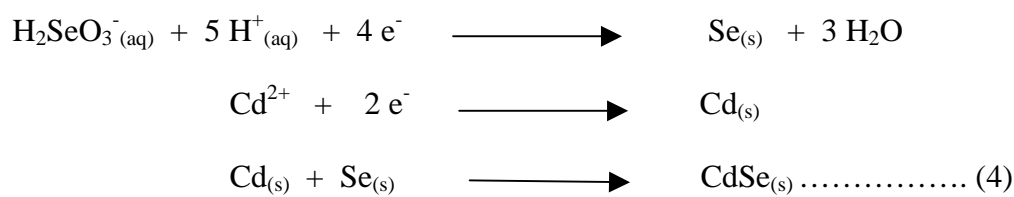

$$H_2SeO_3^-{}_{(aq)} + 5H^+{}_{(aq)} + 4e^- \longrightarrow Se_{(s)} + 3H_2O$$

$$Cd^{2+} + 2e^- \longrightarrow Cd_{(s)}$$

$$Cd_{(s)} + Se_{(s)} \longrightarrow CdSe_{(s)}$$ ……………. (4)

The figure 1 (d) shows the similar reduction current peak at – 0.85 V vs SCE representing the deposition of complex bath at this potential. The two peaks observed at –0.25 V and –0.475 V vs SCE in combined bath corresponds to the reduction of $Se^{2+}$ and $Cd^{2+}$, respectively. The complexing reduction potential is higher than the individual deposition potential. Some workers observed that the reduction potential of CdSe lie between the individual reduction potential of Cadmium and Selenide [19]. Hence the simultaneous deposition of CdSe can be obtained at any potential from –0.475 V vs SCE to –0.85 V vs SCE and can be considered as the underpotential deposition. The depositions carried out at over-potential, -0.85 V vs. SCE, would be diffusion controlled.

Figure 2 shows the polarization curve for the deposition of CdSe onto the Ag/HgBaCaCuO recorded by applying the pulsed potential of 25 Hz frequency and 50 % duty cycle. It shows that the ionic current and hence the deposition of CdSe takes place at about –0.73 V vs SCE.

We have already showed that by applying the sufficient high over-potential, the foreign species can be intercalated into the bulk of the substrate film. Almost all of the electrodeposition reactions are carried out at higher over-potentials and are instantaneous diffusive or progressive diffusive reactions [20]. In the present investigation, we want to avoid the diffusion of the overlayer of CdSe into the HgBaCaCuO film. Hence, it was planed to deposit the CdSe onto HgBaCaCuO at about equilibrium reduction potential or at under potential regime.

Hence it was planned to deposit the film at –0.7 V vs SCE. The chronoamperometry plot measured using VersaStat-II at –0.7 V vs SCE is presented in figure 3. The steady current density after sudden initial decrease is attributed to the uniform deposition onto HgBaCaCuO film. Hence, the deposition in this potential range is not diffusive.

To reveal this, the observed current-time transient during the deposition of CdSe at –0.7 V vs. SCE is fitted with the theoretical instantaneous and progressive growth by plotting the dimensional less plot of $(I/I_{max})^2$ vs $(t/t_{max})$ as shown in figure 4. The relation between $(t/t_{max})$ and $(I/I_{max})^2$ is represented for the instantaneous and progressive growths [21] as follows,

$$(I/I_{max})^2 = 1.9542\ (t_{max}/t)\ [\ 1\text{-}exp\{\text{-}1.2564(t/t_{max})\}]^2\ \text{---------Instantaneous}$$

$$(I/I_{max})^2 = 1.2254(t_{max}/t)\ [\ 1\text{-}exp\{\text{-}2.3367(t/t_{max})^2\}]^2\ \text{--------- Progressive}$$

It is seen that, initially the observed data is closely matched to instantaneous growth, at this stage there is formation of initial critical nuclei and hence current is maximum. After these first few monolayers, if ionic current is sufficiently high then the formation of new nuclei and the diffusion of incoming ions into pre-deposited layer take place and the instantaneous growth profile continues. However the observed data is then closely matched with the progressive growth representing that the generation of new nuclei is very small and the incoming ions gets deposited onto the pre-deposited layer and diffusion free three dimensional growth is in progress.

Figure 5 shows the deposition current density during the deposition of CdSe onto Ag and onto Ag/HgBaCaCuO at the pulsed potential of –0.7 V vs SCE. The current density for the deposition of CdSe on silver substrate is found to be higher (3.85 mA/cm$^2$) than that of deposition on HgBaCaCuO (3.15 mA/cm$^2$). This decrease in deposition current density is attributed to the decrease in conductivity of the substrate after Hg-1212 deposition onto the Ag substrate. This represents that the electrons entered on the surface of Ag, through outer circuitry have to pass through the bulk of oxide HgBaCaCuO film (2 μm thick), which is relatively high resistive than Ag. Hence there is decrease in current. If the current density on Ag/Hg-1212 would be of the same magnitude as that of on Ag, at the same deposition potential, the possibility of diffusion of CdSe ions into the silver substrate would have been considered. This also reveals that the growth is non-diffusive and steady state regime is more constant than that of Ag representing the progressive three-dimensional growth. Smaller deposition current density requires larger deposition period but it gives uniform and non-porous deposit and hence improves the morphology of the films, which in turn influences, the structure of over layer and its properties.

Figure 6 shows the thickness of CdSe deposition onto Ag and Ag/Hg-1212 at −0.7 V vs SCE for different time periods. The thickness is found to be linearly increasing with deposition time. The films are found to be more uniform after first five minutes. The film thickness of the order of 1.5 μm was attained for 24-min. deposition onto Ag/HgBaCaCuO, where the films are found to be of good quality. The thickness of the film onto silver substrate is found to be higher than on Ag/Hg-1212 and is obvious as observed and discussed in deposition current density.

### 3.2 Structural Characterization of Ag/HgBaCaCuO/CdSe Heterostructures:

Figure 7 shows the X-ray diffraction pattern for the as-deposited and annealed (300 $^o$C) CdSe films onto silver substrate deposited for 20 min duration (2 μm thickness). The diffraction pattern was analyzed by using the standard ASTM data [22] and indexed for hexagonal CdSe structure. It is seen that the polycrystalline CdSe is formed on the Ag substrate. The peak of $2\theta = 25.34^o$ corresponds to the diffraction from (002) plane of hexagonal CdSe phase; the stable structure of CdSe at room temperature is normally the hexagonal Wurtzite [23]. The full width at half-maximum FWHM was recorded for this plane for both the samples. Applying the Scherrer's formula to this FWHM, the crystallite size was calculated to be 16 nm and 19 nm, respectively. Hence, after the annealing the crystallite size is increased by 3 nm. The cell parameters were calculated for the as-deposited and annealed samples and are

$$a = 4.29 \text{ Å}, \quad c = 6.97 \text{ Å} \quad \text{at room temperatures.}$$

$$a = 4.29 \text{ Å}, \quad c = 7.01 \text{ Å} \quad \text{at } 300 \text{ }^{o}\text{C.}$$

Figure 8 (a) shows the X-ray diffraction pattern for the CdSe deposited onto Ag/HgBaCaCuO (Hg-1212) heterostructure. It is seen that the pattern contains the planes of tetragonal Hg-1212 and hexagonal CdSe structures. The peaks corresponding to silver substrate are also marked. The presence of (002) characteristic peak of CdSe along with (102), (110), (103) and (203) planes represents the formation of hexagonal CdSe. Similarly, the presence of (001), (002), (111), (005), (104), (114) and (200) planes corresponds to the tetragonal Hg-1212 phase. The $c$-lattice parameter was calculated for both the systems present in the heterostructure and is 12.66 Å for Hg-1212 and 7.01 Å for CdSe. Full width at half maximum (FWHM) was measured for Hg-1212 and CdSe and using Scherrer formula, the crystallite sizes were calculated to be 24 nm and 16 nm, respectively.

The similar X-ray diffraction pattern (figure 8 (b)) can be seen for the deposition of CdSe onto Ag/Hg-1223 films. The reflection peaks are analyzed with tetragonal Hg-1223 system and hexagonal CdSe. The presence of both the system in the deposit is marked in the XRD pattern with corresponding peaks. The lattice parameter '$c$' calculated for both the system are $c = 15.58$ Å for Hg-1223 and 7.01 Å for CdSe. Epitaxial or oriented growth of CdSe could not have achieved as Ag/HgBaCaCuO itself is polycrystalline and there is relative lattice mismatch between HgBaCaCuO and CdSe systems.

### 3.3  Microstructural Characterizations:

Figure 9 shows the scanning electron micrograph for CdSe as-deposited (a), Ag/HgBaCaCuO/CdSe (top view of the junction) (b) and cross sectional view (c) at HgBaCaCuO/CdSe junction. It is seen that the CdSe deposited onto silver substrate is well covered to the substrate and uniform. The grain looks to be spherical in shape with the average grain size of 2 μm.

Figure 9 (b) is the micrograph at the top of lateral junction between HgBaCaCuO and CdSe. Here, CdSe was deposited onto 1 cm$^2$ area, half of the area on HgBaCaCuO (2 cm$^2$), which was deposited onto the rectangular  (1 cm x 3 cm) silver substrate. The two distinct regions with different deposits can be seen. The CdSe film is non-porous, well covered with the substrate (HgBaCaCuO) but the upper surface looks to be slightly rough. This might be due to the increase in relative roughness of the heterostructure, although the uniform and fine-grained HgBaCaCuO film can also be seen from the SEM.

Figure 9 (c) shows typical SEM measured at cross section of HgBaCaCuO/CdSe junction. The presence of distinct layers of CdSe and HgBaCaCuO can be seen with relatively sharp interface boundary between the two deposits. The two regions were probed with EDAX and confirmed. This revels that CdSe is not diffused into HgBaCaCuO deposits. The CdSe granules are found to be bigger in size as that of Ag substrate. This might be due to the slow deposition rate and progressive three-dimensional deposition process as observed from the deposition current density. The non-diffusive growth that predicted by studying the current-time transient and nucleation and growth mechanism is revealed here. Although in figure 9 (b) the top surface is relatively rough, here in figure 9 (c) the interface consisting of HgBaCaCuO and CdSe is observed to the sharp. The atomic force microscope or high-resolution transmission electron microscopic technique is required to further estimate the interface roughness.

### 3.4 Electrical Transport Measurements:

The CdSe film of 1.5 μm thickness was deposited on 1.5 cm length of the HgBaCaCuO deposited films. HgBaCaCuO films used here are single phase Hg-1212. The four equidistant line contracts were made on the top of the heterostructure by using the air drying silver paint. The change in resistivity of film is measured as a function of temperature. The resistivity is normalized to the values at 300 K and presented in figure 10 (a). The normal state (300 K) resistivity of the heterostructure is 3.6 x $10^3$ Ωcm. It is seen that, the resistivity remains constant for initial fall of 15 K, and then starts gradually increasing and remains flat at after 90 K. This represents that, although the resistivity of the HgBaCaCuO layer decreases with temperature, the resistivity of semiconducting CdSe increases at low temperatures. Hence there is an increase in resistivity after 200 K. The resistivity continues to increase further and remains steady after 80 K. Sudden fall in resistivity was excepted at 116 K, as the transition temperature of Hg-1212 used in this system was at this temperature. However, it is interesting to note that the resistivity of individual CdSe layer deposited onto Ag has resistivity of the order of $10^3$ Ωcm (3.6 x $10^3$ Ωcm) and that of HgBaCaCuO deposited on Ag was 17 Ωcm. Because of large difference in resistivity values and dominance of CdSe, the drop resistivity would not have appeared.

The Ag/HgBaCaCuO/CdSe heterostructure sample is then irradiated with He-Ne laser (λ = 632.8 nm, photon energy E = 1.95 eV and the power 2 mW) for 2 hours. Figure 10 (b) shows the resistivity measurements carried out during the laser irradiations. It is seen that the resistivity is decreased after the laser irradiations, but the nature of variations in resistivity with temperature remains similar to that of without laser irradiated samples. The decrease in resistivity is attributed to increase in charge carriers in the heterostructure during laser irradiation. The CdSe has the direct band gap 1.74 eV and hence the laser source with energy 1.95 eV greater than band gap of CdSe was selected. Hence during the irradiations, the electron-hole pair generates in the CdSe sample individually at the junction of the heterostructure. The built-in-junction potential (n) CdSe and (p) HgBaCaCuO heterojunction the electron-hole pairs generated at interface could not recombine, rather holes are transferred to the superconductor region where they can be further trapped into $CuO_2$ layer and hence there is increase in conductivity.

The steady state variation in the resistivity pattern at about 110 K might be due to the transformation of sandwiched HgBaCaCuO into superconducting state. The residual resistivity may be due to higher contact resistance of semiconducting CdSe at HgBaCaCuO/CdSe junction.

Rao *et al.* [8] have developed InAs film of 3500 Å thickness onto TlBaCaCuO film using molecular beam epitaxy technique. They have performed the four point measurements of resistance (R) as a function of temperature (T) using press contacts and found that under-lying TlBaCaCuO is superconducting at 100 K. Whereas the TlBaCaCuO film without InAs was superconducting at 106 K and $T_c$ is not significantly altered by the deposition of InAs. They attributed that the superconducting behaviour was achieved because of the low resistivity InAs as it was produced with the higher doping levels.

Masao Nakao [7] has observed the problem during measurement of resistivity of MgO/BSCCO/GaAs heterostructures. Hence, he has made slight change in geometry of the deposition and deposited GaAs on the middle of the BSCCO films. The contacts were made on the BSCCO on the two sides of GaAs deposits. He observed the enhancement in superconducting transition from 71 K to 83 K by the deposition of GaAs on top of BSCCO. The increase in superconducting parameters was attributed to the fact that GaAs deposited on BSCCO prevents the loss of oxygen from BSCCO film during annealing.

In the present investigation, it was planned to implement the contact geometry as given by Masao Nakao with some slight modifications in current electrodes so that it would be possible to collect all the carriers generated in planar junction during the photo-irradiations.

CdSe was deposited onto the HgBaCaCuO and contacts made are shown figure 11. The HgBaCaCuO films were deposited on Ag substrate with the dimensions of 1.5 cm x 3 cm. The HgBaCaCuO films were deposited with the dimensions of 1.5 cm x 1.5 cm onto silver substrate of dimensions 3 cm x 1.5 cm. The four line contacts were made along the edges of the HgBaCaCuO films as shown in figure 11, leaving sufficient area of 0.8 cm x 1 cm for the deposition of CdSe. The sample area, except middle 0.8 cm x 1 cm was physically covered by the tape and CdSe is then deposited for the optimized pulse-electrochemical parameters.

Figure 12 (a) shows the change in the normalized resistance of HgBaCaCuO as a function of temperature without CdSe depositions. Sample shows zero resistivity at 115 K. The $J_c$ measured at 77 K is 1.7 x $10^3$ A/cm$^2$. Figure 12 (b) show the resistivity of CdSe deposited onto the HgBaCaCuO. It shows the similar resistivity behavior but zero resistivity is achieved at 117.2 K, a slight improvement in $T_c$ is resulted. The $J_c$ values measured to be 1.91 x $10^3$ A/cm$^2$ at 77 K.

This increase in superconducting parameters is attributed to micro-level dislocations induced at the interface of HgBaCaCuO/CdSe heterostructure due to the lattice mismatch between the HgBaCaCuO and CdSe [24]. It is further supported by the fact that some of the 'Se' in CdSe is present in elemental form and is very sensitive to oxygen and forms SeO$_2$. The probability of the presence of extra oxygen species onto the surface of electrochemically oxidized HgBaCaCuO films is already discussed by us [16]. When CdSe is deposited onto the HgBaCaCuO oxide films, the elemental Se forms a bound state with the oxygen present on the surface and the grain boundaries of HgBaCaCuO. Hence, HgBaCaCuO surface at HgBaCaCuO/CdSe junction is physically said to be clean for oxygen inhomogenity; and interface acts as the pathway to carry more supercurrent. But due to bound states of oxygen and CdSe, due to formation of SeO$_2$, the probability of formation of diffusive layer may take place. This type of chemi-sorption of oxygen by elemental Se is discussed by Nair *et al.* [25], but the presence of SeO$_2$ was difficult to identify in X-ray diffraction patterns of nanocrystalline CdSe.

The red He-Ne laser (with $\lambda$ = 632.8 nm, E = 1.95 eV and power P = 2 mW) was irradiated onto CdSe surface of Ag/HgBaCaCuO/CdSe heterostructure, for the 3 hours. The variation in the normalized resistance during laser irradiation was measured as a function of temperature and is shown in figure 12 (c). It can be excitingly observed that the $T_c$ is further increased to 120.3 K and $J_c$ value measured at 77 K is 3.7 x $10^3$ A/cm$^2$. Here, the increase superconducting parameters can only be attributed to the increase in carrier concentration when the sample was irradiated by laser having the energy greater than the band gap of semiconductor.

The minimum energy gap of 1.74 eV of bulk CdSe suggests an onset of optical absorption at 714 nm. The as-prepared film shows the onset of optical absorption near 620 nm and hence an optical band gap is greater than 1.74 eV. Such increase in band gap of chemically deposited CdSe thin films, upto 0.5 eV higher than in single crystal samples has been reported before [25]. This was explained in terms of a quantum size effect arising due to very small crystallite size of ~ 5 nm of the CdSe films. However, in our case the crystallite size calculated by using Scherrer's formula is 16 nm and hence the possibilities of high gap is ruled out. However, we have irradiated the CdSe by red laser having photon energy = 1.95 eV, sufficiently greater than 1.74 eV. The presence of oxygen is notable in the surface layer- a common future of metal chalcogenide thin films. The presence of adsorbed/chemi-sorbed oxygen in the intergrain region improves the photosensitivity of the films.

Hence, the $SeO_2$ at the interface between super-semi heterojunction acts as the trapping centers and the barrier potentials at interface allow the hole to transfer into superconductor and could not be annihilated. Hence, in this way the hole concentration in the superconductor may increase. This might be the reason of the increase in superconducting parameters, particularly the $J_c$ values.

Recently, Foget $et$ $al.$ [26] have discussed the dislocation-induced superconductivity in novel superconducting-semiconducting superlattices (SL). The great body of the experimental data showed that the $T_c$ of the artificial multiplayer as rule does not exceed the transition temperature of the superconducting films constituting SL [27]. As a single exception the semiconducting SL's PbTe/PbS and PbTe/SnTe revealing $T_c$ upto 6 K [28] may be considered. The regular grids of the misfit dislocations on the interfaces of the epitaxial PbTe/PbS SL's are regarded as a phenomenon related to the superconductivity.

There is another experimental fact attracting attention to the dislocations as a phenomenon related to superconductivity. In earlier studies of PbTe/PbS symmetric SL's it was shown that superconducting layers in this system are continued at the interfaces between two semiconductors [29]. On the other hand, in the intrinsically superconducting materials the dislocation strain fields may lead to the formation of localized superconducting domains, surrounding the dislocation at a temperature higher than the 'bulk' transition temperature ($T_c^{zero}$) [30]. At large dislocation densities the superconducting domain may be coupled to a proximity effect, giving rise to global enhancement of $T_c$.

Hence in general, in the present investigation, the increase in superconducting parameters after deposition of CdSe and laser irradiation is attributed to the microlevel dislocations due to lattice mismatch, attachment of selenium to the oxygen present on the surface, development of a diffusion region in semiconductor side of the interface. Upon irradiation, these $SeO_2$ particle acts as trapping centers and the photogenerated holes are transferred to the superconductor and could not be recombined due to the potential barrier of the interface. The holes transferred to the superconductors may be trapped into $CuO_2$ layers and due to increase in hole concentrations the superconducting parameters might have increased.

## 4 Conclusions :

1. The Ag/Hg-based superconductor/CdSe heterostructures are successfully fabricated by a novel pulse-electrodeposition route. This technique works at room temperature and as reactions take place on atomic level; the diffusion of the layers during the growth and processing was avoided.

2. The presence of polycrystalline tetragonal Hg-based phases and hexagonal CdSe phases showed that CdSe could be deposited onto Hg-based films. It is revealed by scanning electron micrograph, where two distinct layers are seen.

3. Apart from the fabrication of device quality Ag/HgBaCaCuO/CdSe heterostructure, deposition of CdSe onto Ag/HgBaCaCuO improves the superconducting parameters of HgBaCaCuO. This is due to the formation of interface states, where the excess oxygen from the surface and boundaries is associated with Se ($SeO_2$) from CdSe, which thereby act as a pathway to improve $T_c$ and $J_c$ values.

4. Further enhancement in $T_c$ after laser irradiation is due to an increase in carrier concentration at the interface where $SeO_2$ formed at interface act as trapping centers and transfer holes to HgBaCaCuO, which thereby increases the superconducting parameters.

**Acknowledgment**

Authors wish to thanks UGC Superconductivity R and D Project and UGC-DRS (SAP) programme for the financial support and Dr. A.V. Narlikar for his constant encouragement. D. D. Shivagan thanks Council of Scientific and Industrial Research (CSIR), New Delhi for the award of Senior Research Fellowship.

.

**Figure Captions :**

Figure 1      LSV recorded for (a) deposition of Cd from 50 mM $CdSO_4$,
         (b) deposition of Se from 10mM $SeO_2$ solutions; and for deposition of CdSe
         (from 50mM $CdSO_4$+10mM $SeO_2$) onto
         (c) Ag, (d) Ag/HgBaCaCuO as a substrate.

Figure 2      Cathodic polarization curve for pulse-electrodeposition of CdSe onto
         HgBaCaCuO thin film at 25 Hz and 50 % duty cycle.

Figure 3      Variation in deposition current density with time during the deposition of
         CdSe onto Ag/HgBaCaCuO at – 0.7 V vs. SCE

Figure 4      Dimension less $(I/I_{max})^2$ vs $(t/t_{max})$ current time transient plot at – 0.7 V vs.
         SCE during the growth of CdSe fitted with the theoretical curves.

Figure 5      Variation in current density during the deposition of CdSe onto Ag and
         Ag/HgBaCaCuO film as substrates.

Figure 6      Variation in thickness of CdSe deposited onto Ag and Ag/HgBaCaCuO at
         different deposition time.

Figure 7      XRD patterns of CdSe films (a) as-deposited and (b) annealed at 300 $^o$C.

Figure 8      XRD patterns for :     (a) Ag/Hg-1212/CdSe heterostructure and
                               (b) Ag/Hg-1223/CdSe heterostructure

Figure 9      Scanning electron micrographs of (a) CdSe film;
         (b) top view of Ag/HgBaCaCuO/CdSe heterostructure; and
         (c) cross sectional view at HgBaCaCuO/CdSe interface.

Figure 10     Variation in normalized resistance of Ag/HgBaCaCuO/CdSe heterostructure
         (a) in dark and (b) under laser irradiations.

Figure 11     The geometry of heterostructure and contacts made for
         the resistivity measurements of HgBaCaCuO films in Ag/HgBaCaCuO/CdSe
         heterostructure.

Figure 12     The change in normalized resistance as a function of temperature for
         (a) Ag/HgBaCaCuO heterostructure
         (b) Ag/HgBaCaCuO/CdSe heterostructure
         (c) Ag/HgBaCaCuO/CdSe during laser irradiations

*Table 1. List of semiconductors with the % lattice mismatch with HgBaCaCuO superconductors*

| Semiconductor | Lattice parameter '$a$' Å | % Lattice mismatch with HgBaCaCuO $a = 3.85$ Å |
|---|---|---|
| Si | 5.43 | 34.06 |
| Ge | 5.65 | 37.89 |
| GaAs | 5.65 | 37.89 |
| InAs | 6.06 | 44.60 |
| CdSe | 4.20 | 8.69 |
| CdTe | 4.57 | 17.10 |
| ZnTe | 6.10 | 31.84 |

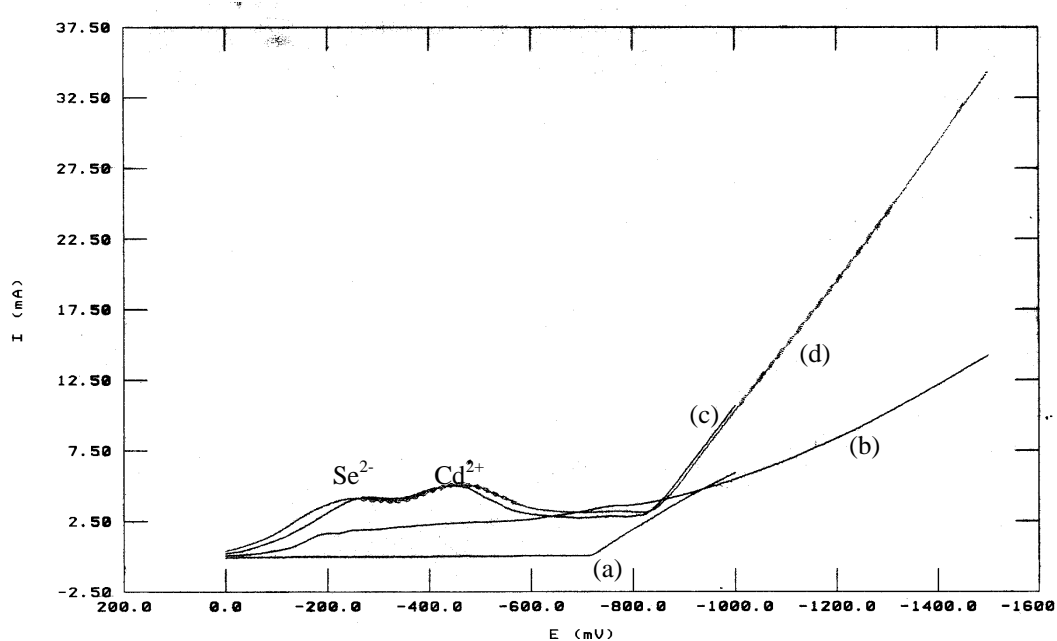

Figure 1      LSV recorded for (a) deposition of Cd from 50 mM $CdSO_4$,
(b) deposition of Se from 10mM $SeO_2$ solutions; and for deposition of
CdSe (from 50mM $CdSO_4$+10mM $SeO_2$) onto
(c) Ag, (d) Ag/HgBaCaCuO as a substrate.

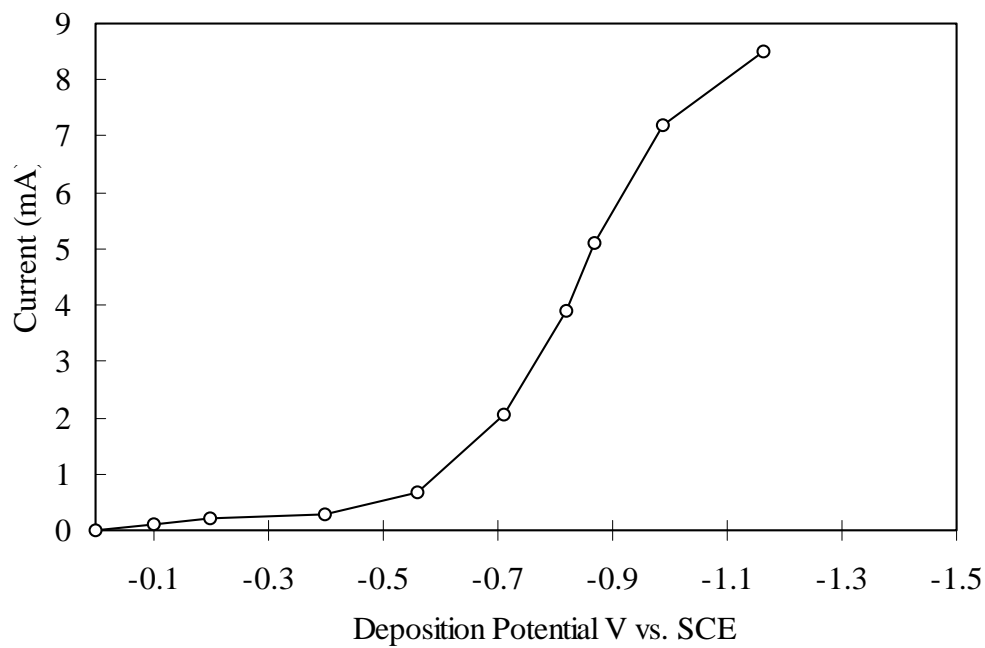

Figure 2      Cathodic polarization curve for pulse-electrodeposition of CdSe onto
HgBaCaCuO thin film at 25 Hz and 50 % duty cycle.

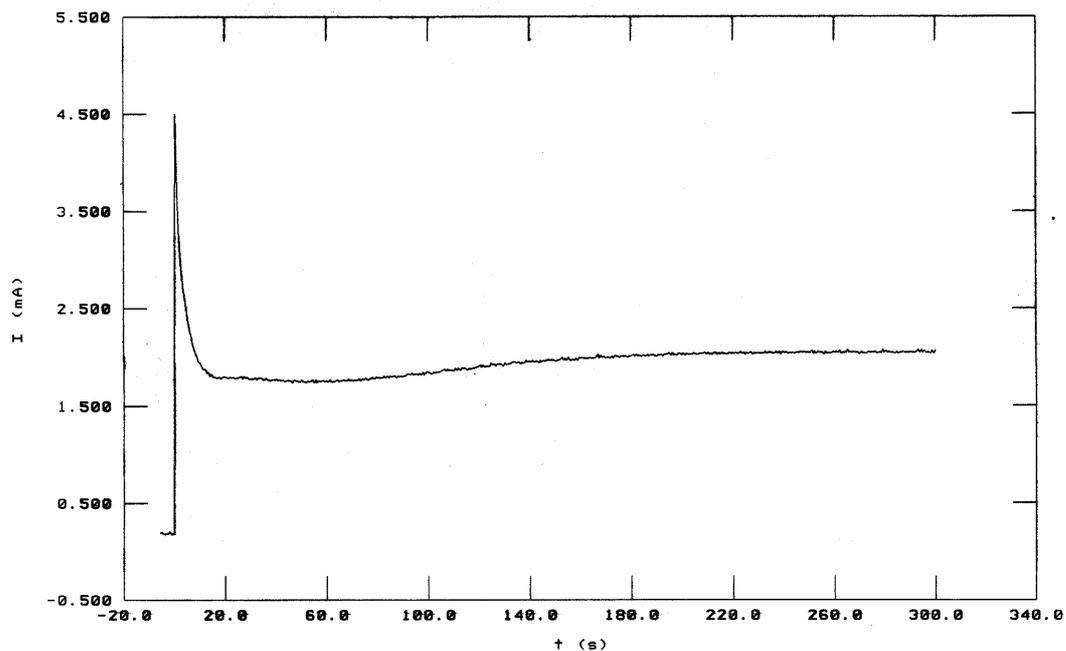

Figure 3      Variation in deposition current density with time during the deposition of CdSe onto Ag/HgBaCaCuO at – 0.7 V vs. SCE

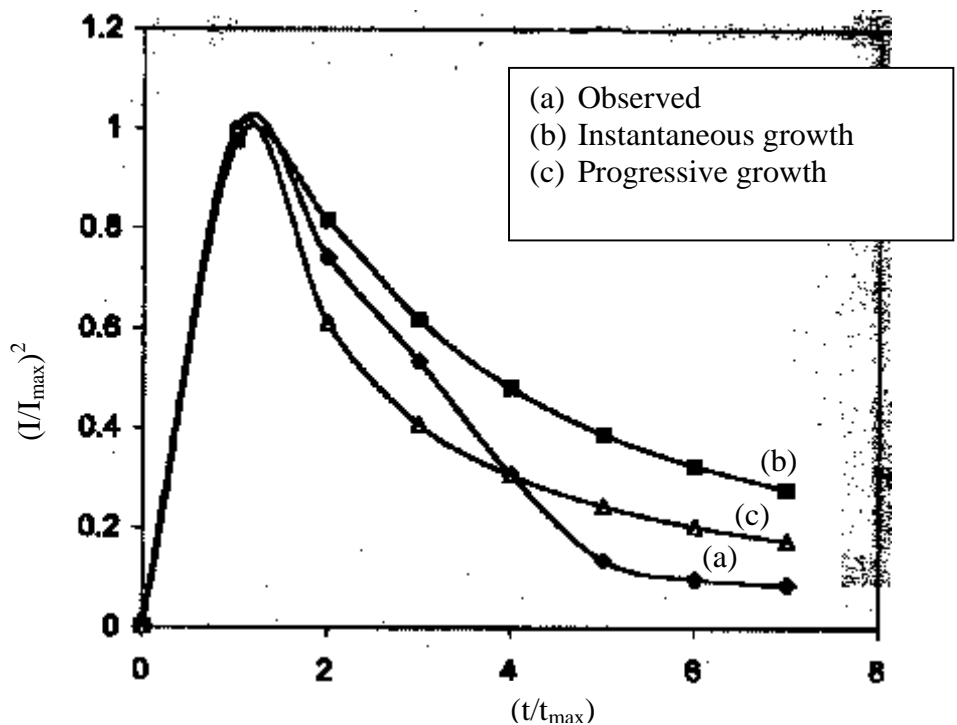

Figure 4      Current time transient at – 0.7 V vs. SCE during the growth of CdSe fitted with the theoretical curves.

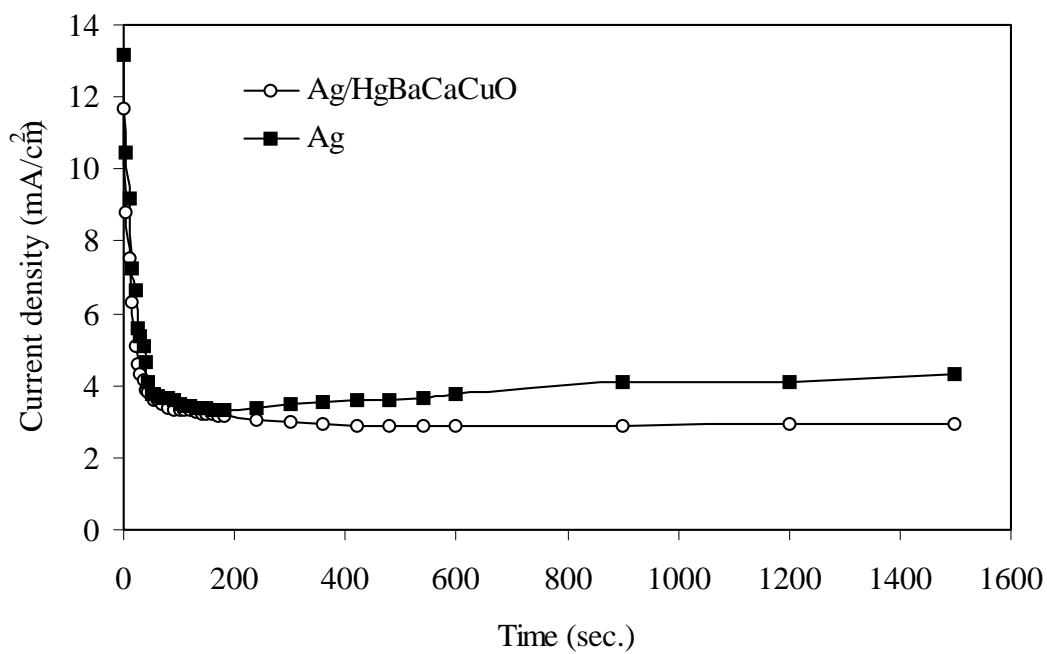

Figure 5        Variation in current density during the deposition of CdSe onto Ag and Ag/HgBaCaCuO film as substrates.

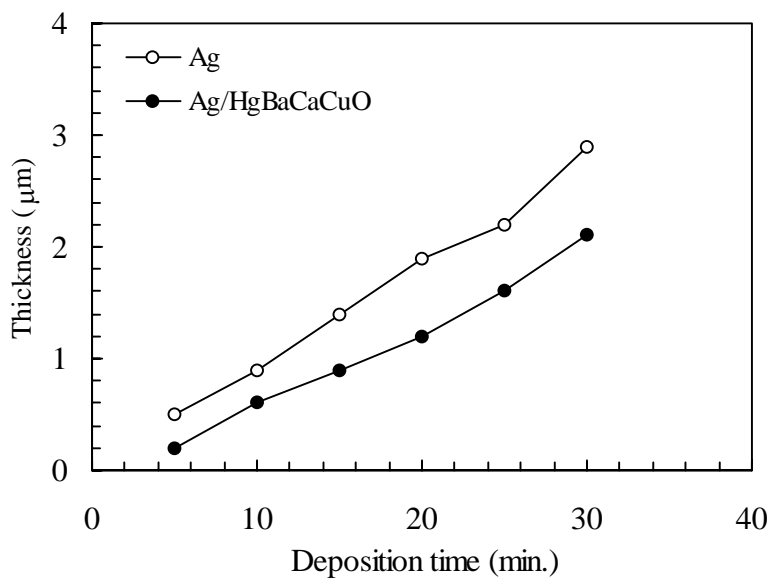

Figure 6        Variation in thickness of CdSe deposited onto Ag and Ag/HgBaCaCuO at different deposition time.

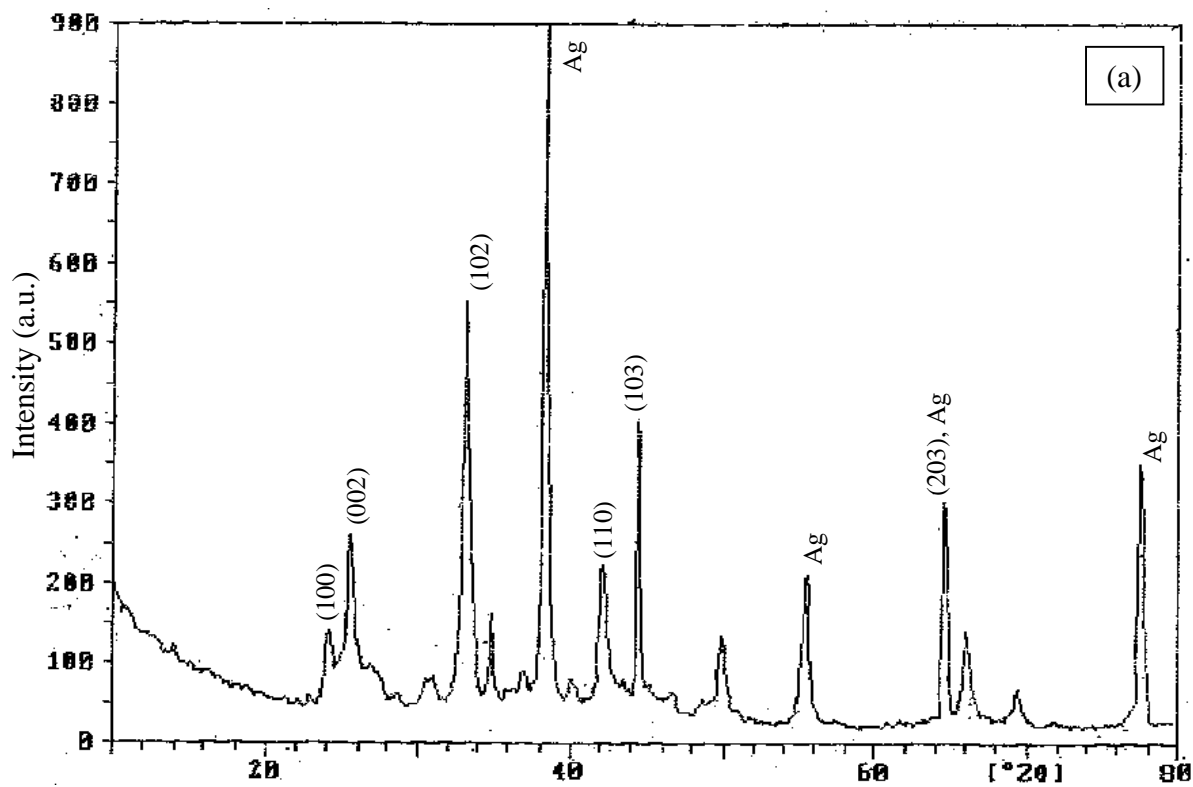

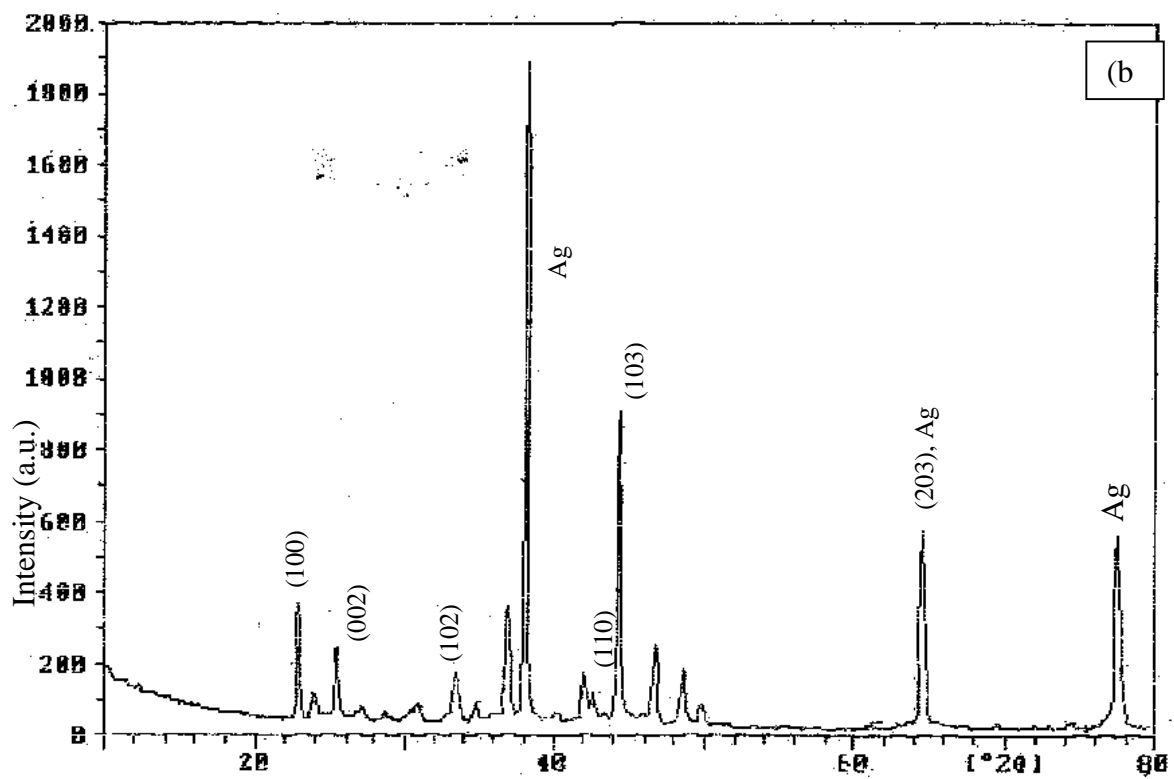

Figure 7 XRD patterns of CdSe films (a) as-deposited and (b) annealed at 300 °C.

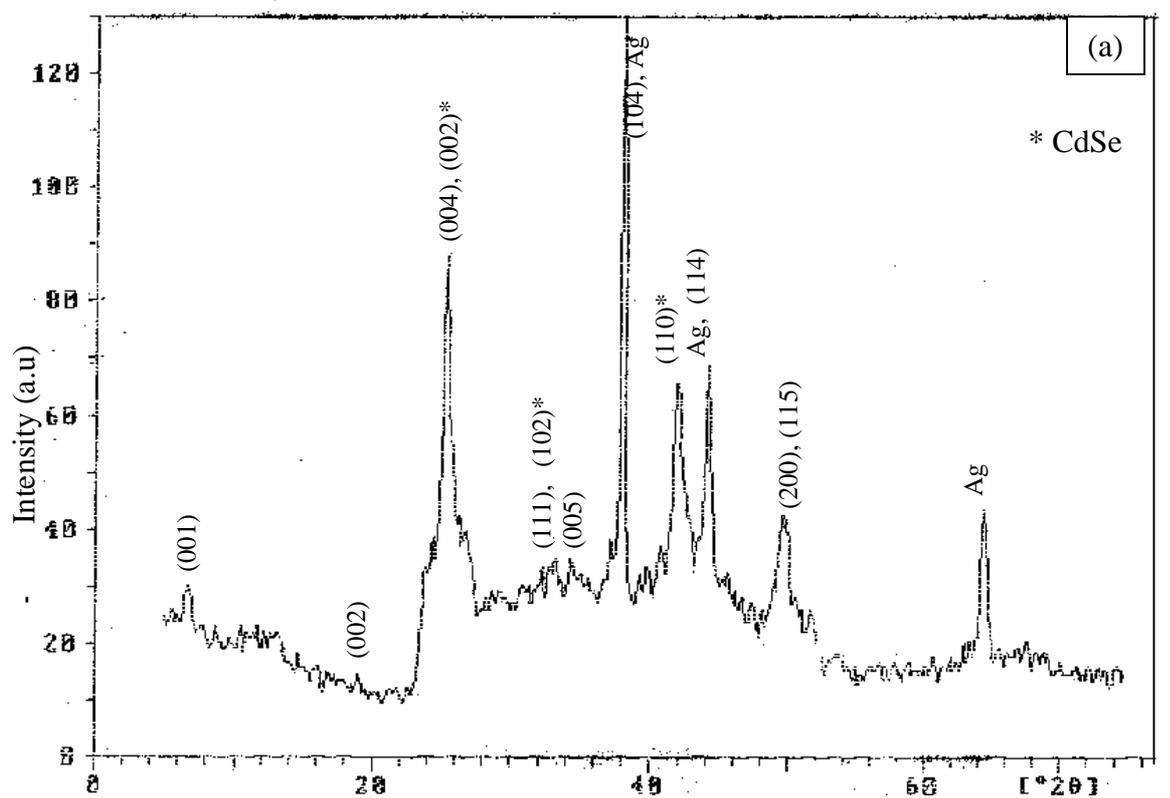

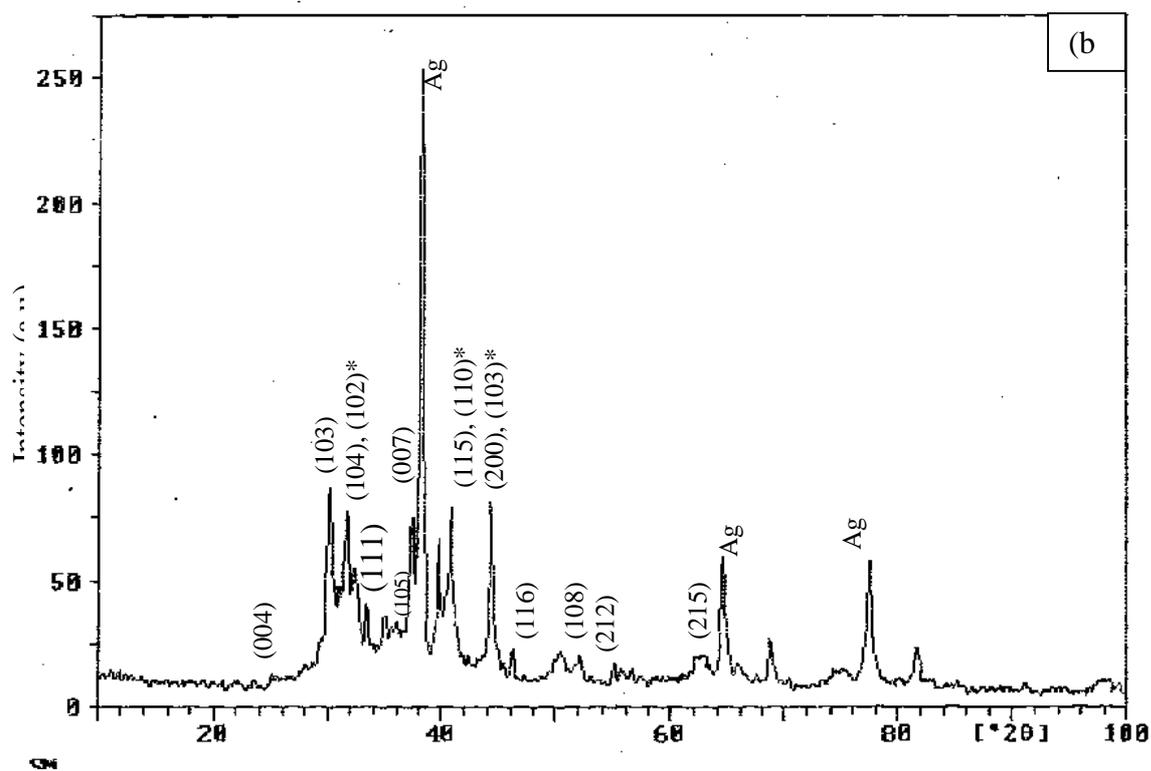

Figure 8     XRD patterns of :     (a) Ag/Hg-1212/CdSe heterostructure and
                                    (b) Ag/Hg-1223/CdSe heterostructure

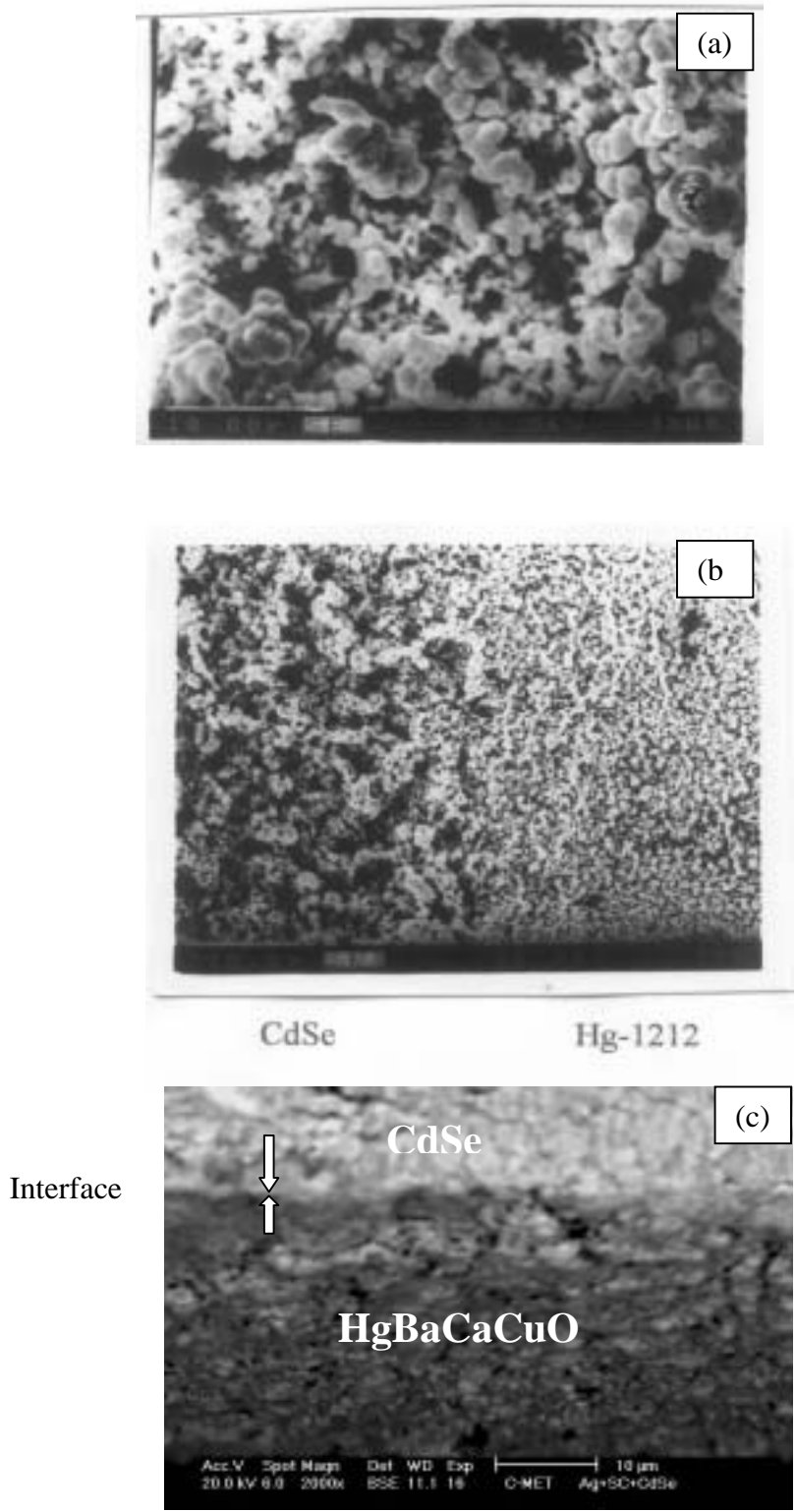

Figure 9     Scanning electron micrographs of a) CdSe film;
            b) top view of Ag/HgBaCaCuO/CdSe heterostructure; and
            c) cross sectional view at HgBaCaCuO/CdSe interface.

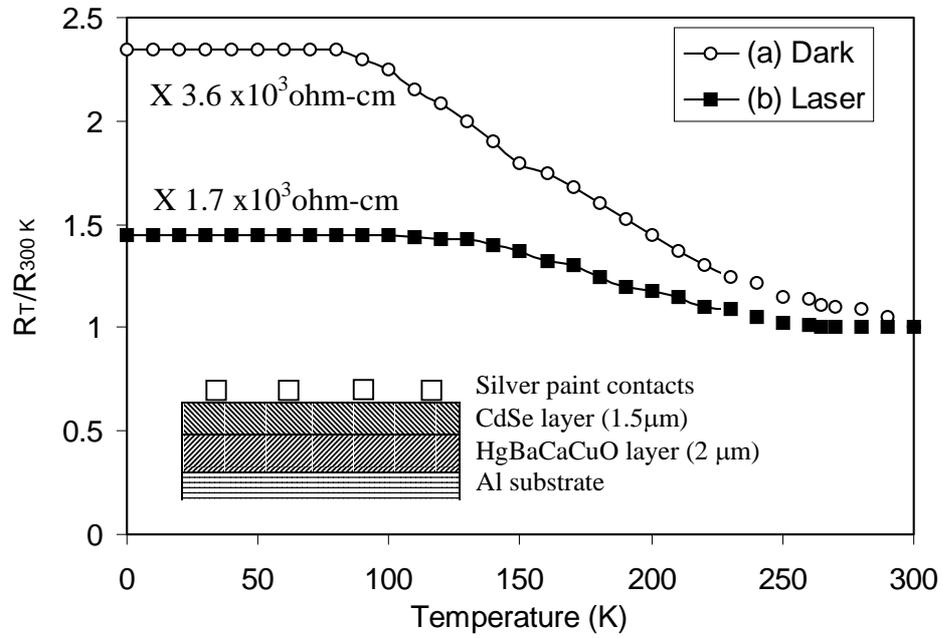

Figure 10. Variation in normalized resistance of Ag/HgBaCaCuO/CdSe
heterostructure in (a) dark; and (b) under laser irradiation.

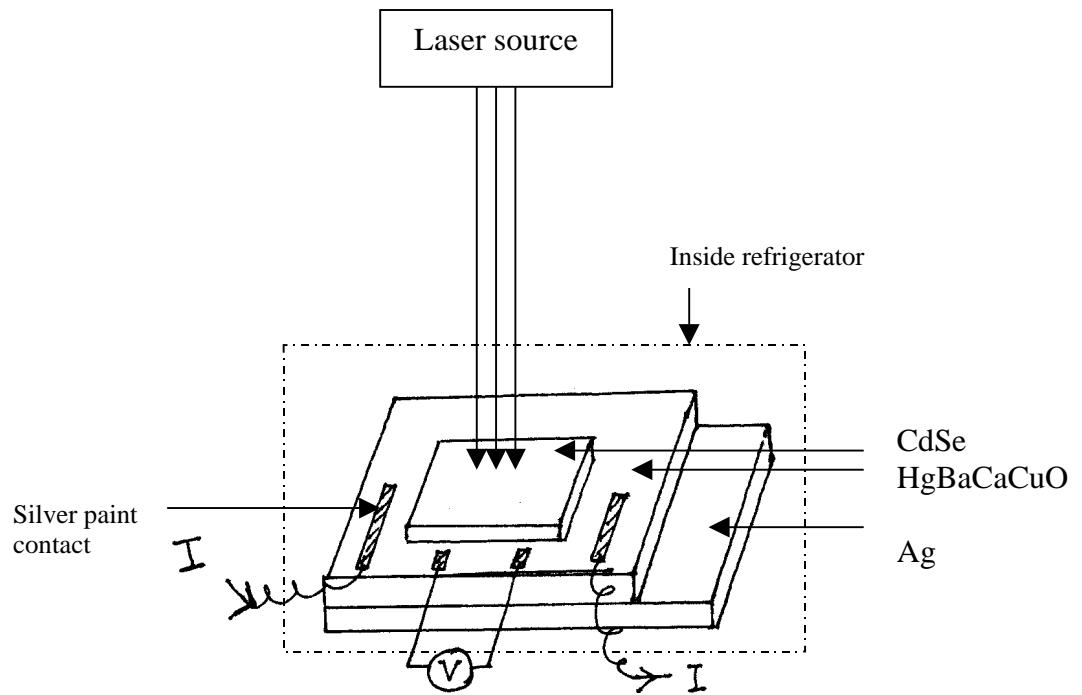

Figure 11     The geometry of heterostructure and contacts made for
the resistivity measurements of HgBaCaCuO films **in**
Ag/HgBaCaCuO/CdSe heterostructure.

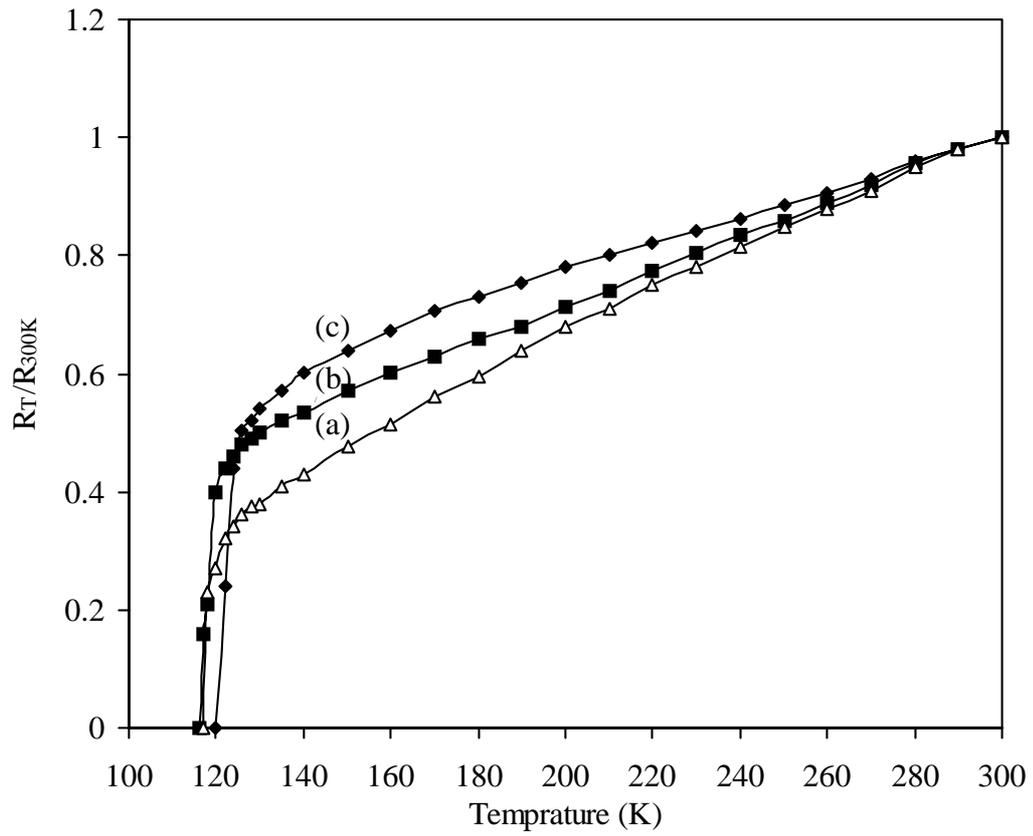

Figure 12  The change in normalized resistance as a function of temperature for
    (a)  Ag/HgBaCaCuO heterostructure
    (b)  Ag/HgBaCaCuO/CdSe heterostructure
    (c)  Ag/HgBaCaCuO/CdSe during laser irradiation.